\documentstyle[12pt,aaspp4,epsf]{article}
\begin{document}
\title{Modeling Gravitational Clustering without Computing Gravitational 
Force} 

\author{Sergei F. Shandarin}
\affil{Department of Physics and Astronomy, University of Kansas, 
Lawrence, KS 66045}
\and
\author{B. S. Sathyaprakash}
\affil{Department of Physics and Astronomy, UWCC, P.O.~BOX 913, Cardiff, 
CF2 3YB, UK}

\begin{abstract}
The large-scale
structure in the Universe is believed to arise out of
small random density perturbations 
generated in the very early Universe, 
that are amplified by gravity.
Large and usually intricate N-body simulations 
are typically employed to model the complex nonlinear dynamics
in a self gravitating medium.
We suggest a very simple model which predicts, on large scales,
the correct density and velocity distributions. 
The model does not involve an explicit computation of the gravitational force.
It is based on a simple transformation of the
variables and local conservation laws of mass and
{\it generalized} linear momentum.
The model demonstrates that the overall appearance of 
large-scale structure in the Universe is
explicitly determined by the initial velocity field and
it reveals the most significant large-scale effects of 
gravity on the formation of structure. 
\end{abstract}

\section{Introduction}
Understanding the formation of the large-scale structure (LSS) in the Universe
is one of
the most important problems in modern cosmology (see e.g. \cite{z-nov83}, 
\cite{pee93}). 
On scales of galaxies complex nonlinear physical processes like turbulence, 
cooling and 
heating of the gas, star formation and supernova explosions, determine
the shape and other properties of objects. However, 
on the scale  of 
superclusters these processes become considerably less important 
and it is reasonable to assume that primarily gravity
shapes the structures. In addition, a variety of observational
evidence supports the hypothesis that most of the mass in the Universe
is in the form of weakly interacting particles that feel only gravity.
Therefore, a simple physical model  assuming that the Universe is 
filled with a cold
dust-like gas only interacting gravitationally 
is most often used for studying the LSS formation. 
The baryonic component plays no crucial role except on the scale of galaxies. 

When the amplitude of the density fluctuations is small, their growth is
adequately described by the linear theory. 
However, as soon as the amplitude
of the perturbations reaches the order of unity 
($\sigma_{\delta} \equiv\left <[(\rho - \bar{\rho})/\bar{\rho}]^2 
\right>^{1/2} \approx 1$, where $\rho$
is the density field and $\bar \rho$ its mean) the linear 
theory breaks down (see e.g. \cite{z-nov83}, \cite{sh-z89}). 
The most widespread method to deal with the complex dynamics at the
nonlinear stage is to run N-body simulations generating 
the initial condition
as a realization of a Gaussian random process (\cite{kly-sh83}). 
In N-body simulations of this type 
the gravitational
forces generated by the density distribution is calculated 
at each time step. The trajectory
of every particle is integrated in a self-consistently varying 
gravitational field. Cosmological N-body simulations have played the
most significant role in testing (and in most cases rejecting) the
models for dark matter. Here we are interested in a different aspect
of the problem of the LSS formation. We are trying to formulate simple
physical {\it macroscopic} principles controlling the nonlinear stage of
gravitational clustering. 

Long ago Zel'dovich (1970)  suggested a very elegant, analytical 
approximation to describe the beginning of the nonlinear stage in
cosmological scenarios assuming smooth initial conditions.
Quantitatively it can be expressed as a requirement that the initial 
power spectrum of density fluctuations  has a steep cutoff on small scales
(steeper than $P(k) \propto k^{-3}$).
Mathematically, the Zel'dovich Approximation (ZA) is a one step mapping from
the  Lagrangian space into the Eulerian space at a time $t,$ 
given by
 \begin{equation}
 {\bf r}({\bf q},t)=a(t)[{\bf q} -D(t) \nabla \Phi({\bf q})]
 \end{equation}
 where ${\bf r}$ and   ${\bf q}$ are the Eulerian and Lagrangian coordinates,
respectively;
 $a(t)$ is the scale factor describing the 
homogeneous expansion of the Universe;
 $D(t)$ is a known function of time describing the growth of perturbations;
 and $\Phi({\bf q})$ is the potential of the initial velocity field:
${\bf v}_0 \propto -\nabla \Phi({\bf q})$. 
With the aid of the above mapping one can
calculate the density at the final time $t$ using the conservation of mass.  
However, cosmological observations almost certainly exclude scenarios having 
no perturbations on small scales.
Small scale power is required to explain the existence of 
quasars and galaxies at very
high redshifts. 

Two modifications---the Truncated 
Zel'dovich Approximation (TZA) (\cite{kof-etal92}, \cite{Col-etal93}) 
and the Adhesion Approximation (AA) (\cite{gur-sai-sh89})---have 
been suggested in an attempt to extend the scope of the ZA and to make it
more useful for general cosmological scenarios
(for a brief review see e.g.~\cite{sh94} and for more exhaustive discussion
of various approximations see \cite{sathya-etal95}, \cite{sah-col95}). 
A comparison with gravitational N-body simulations shows that these
two approximations fairly accurately describe nonlinear gravitational
clustering (\cite{mel-sh-wei94}, \cite{sathya-etal95}).
Here we describe another approximation to the nonlinear 
gravitational evolution
of the density and velocity perturbations which is numerically almost as simple
as the ZA but completely universal and much
superior to any approximation method suggested so far. It consists of two 
elements: 
\begin {itemize}
\item a transformation of variables and
\item an assumption that in the
dense regions the local gravitational interaction can be approximated by
the diffusion of a generalized momentum (see below for a definition of
the generalized momentum).
\end {itemize}
The latter assumption is similar to
the AA but in this model the generalized momentum  
is locally conserved.

\section{COnserving Momentum Approximation (COMA)}
It is well known that one can exclude many effects of 
the homogeneous and isotropic
expansion of the 
Universe by introducing the comoving coordinate ${\bf x} = {\bf r}/a(t)$
and peculiar velocity ${\bf v}_p = {\bf u}- H(t){\bf r}$ where ${\bf u}$
is the physical velocity and $H(t)=\dot{a}/a$ is the Hubble parameter
characterizing the rate of expansion of the Universe. In addition, it is
convenient to parameterize time by $D(t),$ which describes 
the growth of perturbations, and rescale the density and peculiar 
velocity (\cite{gur-sai-sh89}) by
\begin{equation}
\rho({\bf x},t)=a^{-3} \eta({\bf x}, D(t)) , ~~~~~~
{\bf v}_p({\bf x},t) =(a \dot{D}) {\bf v}({\bf x}, D(t)).
\end{equation}
In terms of the new variables the dynamics of gravitational clustering
is described by (i) the conservation of mass, (ii) a dynamical equation
analogous to the Euler equation 
\begin{equation}
 {{\partial v_i} \over {\partial D}} + v_k {{\partial v_i} 
 \over {\partial x_k}}
 = {3 \over 2} {\Omega_0 \over {D \cdot ({d \ln D \over d \ln a})^2}} 
({\partial \varphi \over \partial x_i} + v_i), \label{eq:euler-D}              \end{equation}
where $ \Omega_0=8 \pi G \bar{\rho}/ 3 H_0^2 $ is the dimensionless 
mean density of the Universe, and (iii) an equation for the gravitational 
field $\varphi$ analogous to 
the Poisson equation (for derivations see e.g. \cite{sh94}).

As in the ZA the COMA assumes that the
gravitational force is approximated by the velocity 
\begin{equation}
{\partial \varphi /
 \partial x_i}  \approx - v_i({\bf x},D) 
\end{equation}
which sets the right hand side of
eq.[\ref{eq:euler-D}] to zero. Thus, on large scales the model mimics 
the dynamics of a self-gravitating
medium by the effectively kinematic model of non-interacting medium.

However, on small scales the approximation  assumes 
that particles inelastically collide resulting in the exchange and diffusion
of the generalized momentum defined as
\begin{equation}
{\bf p} \equiv \eta {\bf v} = {a^2 \over \dot{D}} \rho {\bf v}_p,
\end{equation}
which mimics the effects of gravity in dense regions (\cite{sh-z89}).

In the AA the right hand side of eq.[\ref{eq:euler-D}] is supplemented 
by the viscosity term resulting in  Burgers' equation
\begin{equation}
{{ \partial v_i} \over {\partial D}} +
 v_k {{ \partial v_i} \over {\partial x_k}}
={\nu} ~{\nabla}^2 v_i \label{eq:burg}.
\end{equation}
In the AA $\nu$ is assumed to be constant which results in the violation
of the momentum conservation (\cite{kof-sh90}) and in this respect the
model is not physical. The COMA assumes physical viscosity 
described by the equation similar to the Navier-Stokes equation
(here we ignore the 2nd viscosity)
\begin{equation}
{{ \partial v_i} \over {\partial D}} +
 v_k {{ \partial v_i} \over {\partial x_k}}
={1 \over \eta} {\partial \over \partial x_k}
[\mu({\partial v_i \over \partial x_k}+{\partial v_k \over \partial x_i}
-{2 \over 3}\delta_{ik}{\partial v_l \over \partial x_l})] \label{eq:navier}.
\end{equation}
where $\mu$ is the dynamical viscosity. 
The Navier-Stokes equation conserves the linear momentum automatically
but does not have an analytical solution in the general case; also it
may generate vorticity.

According to the ZA in the single stream flow regime each
particle moves along straight line with constant velocity. However,
as soon as a particle enters a multi-stream flow region its trajectory
becomes very complex, resembling a random walk. Collectively this type 
of motion can be labeled as a sort of violent relaxation (\cite{lin67})
and can be roughly
approximated by the diffusion of the momentum. It is convenient to split
the force exerting on the particle into two parts:
The first component represents
its interaction with the other particles of the clump of which it is a member
and the second  represents its interaction with particles that 
belong to other clumps or voids. 
One can reasonably assume that the irregular component
of the motion is mostly determined by the gravitational force induced by
the particles from the same clump. If so, the 
momentum of the clump must be approximately conserved in the interaction
of this type. 
The change of the momentum of the clump
is due to interaction with the surrounding matter that can be roughly described 
by eq.[4].
In practice, the diffusion of momentum is important in all regions
of high density (filaments and pancakes) and aids in preventing 
the formation of multi-stream flows as in the AA.
The major physical difference between the COMA and AA consists in that
the COMA conserves momentum locally 
and the AA, based on Burgers' equations, does not
(instead, the latter conserves the velocity). We believe that this an
advantage of the COMA.

Another advantage of the COMA is mainly practical. 
The numerical code realizing the model is extremely simple and efficient.
It operates with the density and velocity distributions (no particles!) on 
a cubic grid in the Eulerian space: $\eta = \eta({\bf x},D)$ 
and ${\bf v} = {\bf v}({\bf x},D)$. 
At each time step it calculates the flows of mass and momentum from a
mesh cell to its neighbors and computes the new density and velocity
fields on the grid using a cloud-in-cell algorithm (see e.g. \cite{hoc-eas88})
thereby explicitly
conserving the mass and the generalized momentum (simulation of
perfectly inelastic collisions). 
Thus, one circumvents the problem of having to compute 
the gravitational force after each time step.  
A full description of the code as well as the code itself will
be published elsewhere; here we list its major features.
The algorithm implementing the new approximation is:
\begin{itemize}
\item {\it universal} in terms of the initial and boundary
conditions as well as the shape of the computational box,
provided that the initial velocity field is somehow generated;  
\item {\it extremely simple}  both conceptually and practically;
\item {\it very economical} in terms of memory
(just four arrays, each as big as the size of the box, are sufficient to
implement the algorithm); 
\item {\it very fast}; and
\item local, therefore {\it 100\% parallelizable}. 
\end{itemize}
Generalization of the algorithm to deal
with particles, rather than densities on the grid,
as well as introducing partially inelastic collisions is straightforward 
but would require additional memory and slightly slow down the code.

The only disadvantage of the COMA (compared to the AA) is that it does not
have an analytic solution. 

\section{Comparison with N-body simulations}
In order to test the strength of our approximate model
we compare it with gravitational N-body simulations.
The N-body simulations are performed using $128^3$ particles on a 
$128^3$ mesh with periodic boundary conditions. 
The power spectrum of primordial density fluctuations is assumed to 
be a simple power law ($|\delta_k|^2\propto k^n,$ where $\delta_k$ is the
Fourier transform of the density contrast $\delta$)
covering the range $n= +1,$ $0,$ $-1$ and $-2$.  
The initial fluctuations are evolved gravitationally and comparisons 
are made when different scales, determined by the parameter $k_{nl}$ 
(defined by the equation $\sigma^2_{\delta}=a^2 \int_0^{k_{nl}}P(k)\;d^3k=1$),
go nonlinear:
$k_{nl}= 64, 32, 16, 8$ and $4$ (in units of the fundamental frequency
$k_f=2\pi/L$, where $L$ is the size of the box).
For a detailed discussion of the N-body simulations 
see \cite{mel-sh93}.
At each epoch the density field is computed on a reduced grid
of size 64$^3$ by a cloud-in-cell method. 

We run the model code with identical initial conditions and compare
its results with those of N-body simulations
at different stages. Here we illustrate this comparison for three
different values of the power-law index of the spectrum of fluctuations,
namely,  $n=0,$ $-1$ and $-2,$ and at two stages $k_{nl}=8$ and $4$.
Although it is unlikely that such power-law models can explain the 
real universe, they serve as toy models to understand generic features of 
gravitational instability. The following normalizations provide a rough 
idea of the scales involved and how such toy models may relate to 
the real Universe. 
One can view the density fields corresponding to
different stages in the evolution of such power law density fluctuations
as equivalent to  the
density distribution at the present epoch but obtained 
after smoothing with a top-hat filter of progressively smaller
radii $R_{TH},$ with smaller smoothing lengths corresponding 
to later epochs.  For instance, the epochs $k_{nl}=8$ and $4$ correspond
to smoothing radii 
$R_{TH} \approx 2$, and $1~\,~h^{-1}~Mpc,$ respectively, within
volumes $(200\,h^{-1}~Mpc)^3$, and $(100\,h^{-1}~Mpc)^3.$

\begin {figure} [h!]
\begin {center} \leavevmode \epsfxsize=5.0 in \epsfbox {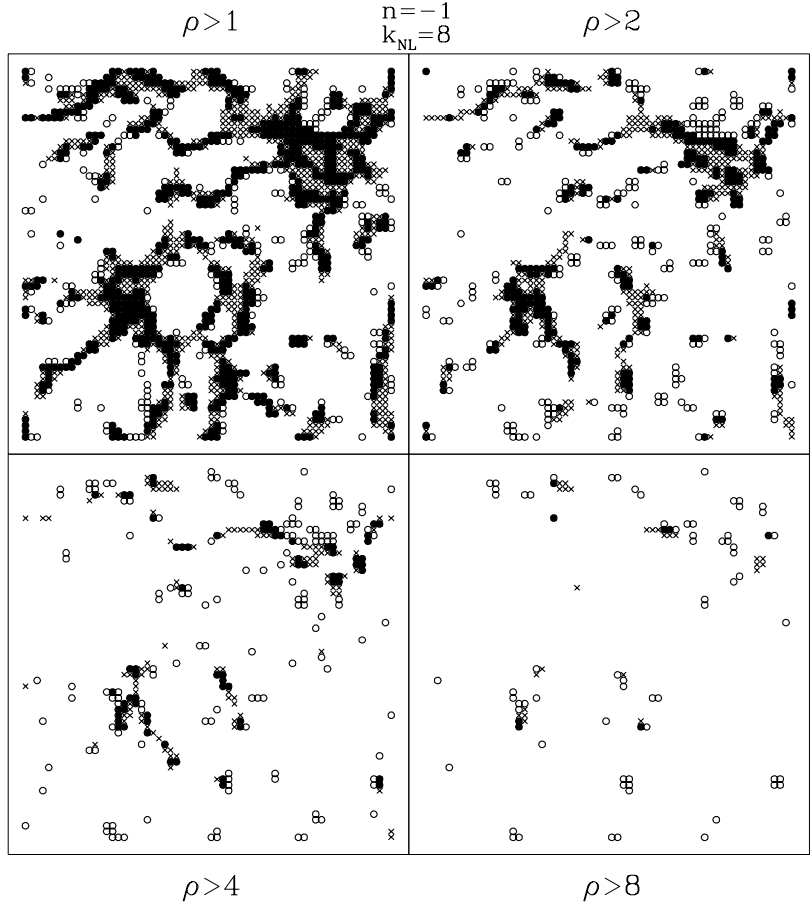} \end {center}
\caption {Sites with densities
exceeding $\rho >1, 2, 4$, and $8$ times the mean density
 in the $n=-1$ model at the $k_{nl}=8$
stage, are shown. Empty circles correspond to sites where only the 
N-body densities are larger than the density threshold, crosses represent
sites where only the COMA densities are larger than the
threshold and filled circles show sites where both
the distributions are larger than the threshold.}
\end {figure}

In Fig. 1 we compare the N-body density distributions (open circles)
with those obtained using the COMA (crosses) in 
a typical two-dimensional slice, one mesh size thick, in the 
$n=-1$ case, when the scale of nonlinearity corresponds to $k_{nl}=8.$
The panels show sites of density contrast ($\delta$)
larger than a certain threshold ($\delta_c$). A filled square
is a site where both the 
N-body and model densities  are larger than the threshold.
One can see that there is generally a good overall agreement 
between the two distributions at nonlinear but not extremely high density
thresholds. However, the COMA 
shows a rather smoother distribution and the density peaks 
that it produces are somewhat lower, 
especially at higher densities.  This behavior is not
unexpected: local nonlinear gravitational effects, ignored by the COMA, 
make the density distribution
more clumpy and the clumps are more compact. 
A detailed comparison shows that the vast majority of density
peaks in the N-body simulations have their counterparts in the COMA 
often shifted by a couple of mesh sites and 
of slightly lower amplitudes.
In order to quantify the similarities and 
discrepancies of the two density distributions in Fig.~2
\begin {figure} [h!]
\begin {center} \leavevmode \epsfxsize=5.0 in \epsfbox{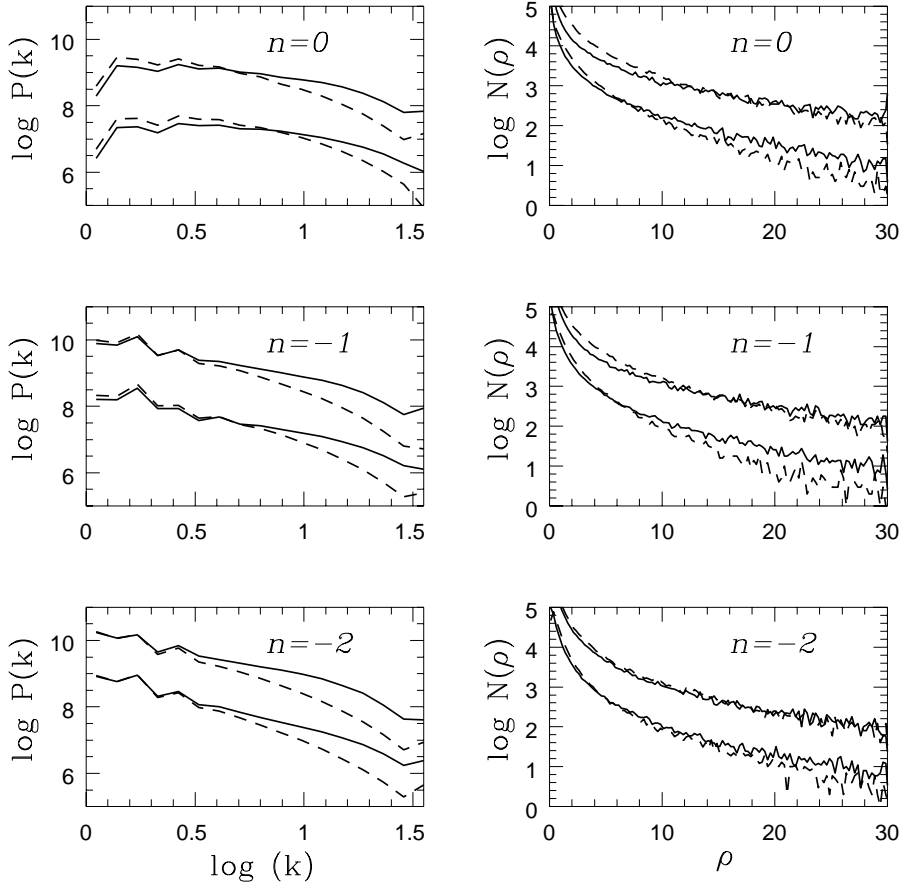} \end {center}
\caption {The left column shows the power spectrum in three models at two
stages each: the bottom curves correspond to the $k_{nl}=8$ and the top
curves correspond to the $k_{nl}=4$. Solid lines correspond to the N-body
simulations and dashed lines to the COMA. Similarly, the right
column shows the density distribution functions normalized to the total
number of mesh sites ($64^3$). In order to avoid overlapping the curves
for the $k_{nl}=4$ stage are multiplied by a factor of 10.}
\end{figure}
we plot the power spectra (left panels) and  probability 
distribution functions (right panels) of the N-body (solid lines)
and the COMA (dashed line) density fields, for three
values of the power-law index
$n=0,$ $-1$ and $-2,$ (panels from top to bottom) and 
two epochs $k_{nl}=8$ (bottom pair of curves) and $4$ (top pair of curves).
Comparisons of the power spectra show that the COMA 
does not have enough power on
small scales which is obviously related to a relative smoothness of the 
density fields in the COMA. Similarly, the density distribution
functions show that when $k_{nl}=8,$ sites with densities 
above $10$ (in units of the mean density) are less abundant in the 
COMA. However, at a later stage when $k_{nl}=4$ the 
COMA density
distribution functions reproduce the N-body distribution
correctly up to densities $\rho = 30$, but at larger densities
(not shown) they again fall down faster than the 
ones in the N-body simulations.  A similar comparison of N-body with
other successful models, such as the AA or the TZA, 
does not reveal as great an agreement as between
COMA and N-body (\cite {mel-sh-wei94}).

\section{Summary}
Summarizing, we conclude that the proposed approximation mimics the 
large-scale gravitational evolution at late nonlinear stages quite well
(positions of clumps, filaments, pancakes) --
better than any other known approximation. 
This implies that the two simple assumptions:
\begin {itemize}
\item the generalized gravitational force on large scales 
($l\ge k_{nl}^{-1}$) equals the velocity :
${\partial \varphi / \partial x_i}  \approx - v_i({\bf x},D)$ and
\item the conservation of mass and 
the local ($l\le k_{nl}^{-1}$) conservation of the generalized momentum, 
${\bf p} = \eta {\bf v}  $, \\
\end {itemize}
explain fairly well the nonlinear gravitational clustering on large scales.

In a very general sense the COMA falls in the class of sticky particle methods
used in the numerical hydrodynamics and also resembles the lattice gas
models used in modeling turbulence and similar phenomena.
An interesting question is whether the COMA guarantees complete hierarchical
clustering or not. Due to numerical errors some small clumps may miss 
merging with the
larger ones passing by without the collision. We have not seen this
phenomena in our simulations and believe that it should not be a serious
problem for the method.

We believe that after a thorough testing the approximation can be a practical
tool for cosmological studies of large-scale processes which do not require
a resolution better than a few $Mpc,$
such as large-scale streaming velocities, spatial distribution of rich
clusters of galaxies, statistics of voids, etc. In such low resolution
calculations the COMA
can be more efficient than N-body simulations if very large volumes and large
statistical ensembles are required.

\acknowledgments

We are grateful to Lev Kofman, 
Dmitri Pogosyan, Varun Sahni, and Capp Yess for discussions.
Acknowledgments are due to the Smithsonian Institution, Washington,
USA, for International travel assistance under the ongoing Indo-US
exchange program at IUCAA, Pune, India.
S. Shandarin acknowledges the financial support from NASA Grant
NAGW-3832, NSF Grant AST-9021414, and the University
of Kansas grant GRF 96.

\end{document}